\documentclass[journal]{IEEEtran}

\usepackage{amsmath}
\usepackage{amsfonts}
\usepackage{amssymb}
\usepackage{siunitx}

\usepackage{graphicx}

\usepackage{booktabs}
\usepackage{makecell}

\usepackage[natbib, bibencoding=utf8, citestyle=numeric, bibstyle=ieee, maxbibnames=15, maxcitenames=2, mincitenames=1, sortcites]{biblatex}
\bibliography{bibliography}

\usepackage{hyperref}
\hypersetup{hidelinks}
\usepackage[capitalise, nameinlink, noabbrev]{cleveref}

\usepackage[acronym, shortcuts, nohypertypes={acronym}]{glossaries}
\newacronym{CCC}{CCC}{concordance correlation coefficient}
\newacronym{SER}{SER}{speech emotion recognition}
\newacronym{PCC}{PCC}{Pearson correlation coefficient}
\newacronym{MAE}{MAE}{mean absolute error}
\newacronym{MDE}{MDE}{mean directional error}
\newacronym{MSE}{MSE}{mean square error}
\newacronym{RMSE}{RMSE}{root mean square error}
\newacronym{DNN}{DNN}{deep neural network}
\newacronym{PPC}{PPC}{precision per class}
\newacronym{RPC}{RPC}{recall per class}
\newacronym{UAP}{UAP}{unweighted average precision}
\newacronym{UAR}{UAR}{unweighted average recall}
\newacronym{SNR}{SNR}{signal-to-noise ratio}
\newacronym{NLP}{NLP}{natural language processing}
\newacronym{AMR}{AMR}{adaptive multi-rate}
\newacronym{spearmans}{Spearmans rho}{Spearman’s rank correlation coefficient}
\newacronym{crema-d}{CREMA-D}{Crowd-sourced Emotional Multimodal Actors Dataset}
\newacronym{des}{DES}{Danish Emotional Speech}
\newacronym{emodb}{EmoDB}{Berlin Database of Emotional Speech}
\newacronym{emovo}{EMOVO}{Italian Emotional Speech Database}
\newacronym{iemocap}{IEMOCAP}{Interactive Emotional Dyadic Motion Capture}
\newacronym{meld}{MELD}{Multimodal EmotionLines Dataset}
\newacronym{msppodcast}{MSP-Podcast}{MSP-Podcast}
\newacronym{pesd}{PESD}{Polish Emotional Speech Database}
\newacronym{ravdess}{RAVDESS}{Ryerson Audio-Visual Database of Emotional Speech and Song}
\newacronym{musan}{MUSAN}{Music, Speech, and Noise Corpus}
\newacronym{nsc}{NSC}{Singapore English National Speech Corpus}
\newacronym{mardy}{MARDY}{Multichannel Acoustic Reverberation Database at York}
\newacronym{air}{AIR}{Aachen Impulse Response}
\newacronym{timit}{TIMIT}{TIMIT Acoustic-Phonetic Continuous Speech Corpus}

\begin{document}

% === title ===
%BS: inserted line break for nicer appearance:
\title{
    Testing Correctness, Fairness, and Robustness \\of Speech Emotion Recognition Models
}
%BS: replaced e.g. by e.\,g., and i.e. by i.\,e., throughout

% === authors ===
\author{
Anna~Derington,
Hagen~Wierstorf,
Ali~\"{O}zkil,
Florian~Eyben,
Felix~Burkhardt,
Bj\"{o}rn W.\ Schuller
\thanks{
e-mail: aderington@audeering.com
}
\thanks{A.\ Derington and H.\ Wierstorf are co-first authors.}
\thanks{
A.\ Derington,
H.\ Wierstorf,
F.\ Eyben,
F.\ Burkhardt,
B.\,W.\ Schuller
are with audEERING GmbH, Gilching, Germany.
}
\thanks{
A. \"{O}zkil
is with Jabra, GN Audio, Copenhagen, Denmark
}
\thanks{
B.\,W.\ Schuller
is, in addition, with CHI -- Chair of Health Informatics, Technical University of Munich, Germany;
GLAM -- Group on Language, Audio, \& Music, Imperial College London, UK.
}
}

\maketitle

\begin{abstract}
Machine learning models for speech emotion recognition (SER)
can be trained for different tasks and are usually evaluated
based on a few available datasets per task.
Tasks could include arousal, valence, dominance, emotional categories, %BS: added:  
or 
tone of voice.
Those models are mainly evaluated in terms of correlation or recall,
and always show some errors in their predictions.
The errors manifest themselves in model 
%BS: unifying to British English throughout...
behaviour,
which can be very different along different dimensions
even if the same recall or correlation is achieved by the model.
This paper introduces a testing framework
to investigate behaviour of speech emotion recognition models,
by requiring different metrics
to reach a certain threshold in order to pass a test.
The test metrics can be grouped 
in terms of correctness, fairness, and robustness.
It also provides a method for automatically specifying test thresholds
for fairness tests,
based on the datasets used,
and recommendations on how to select the remaining test thresholds.
We evaluated a xLSTM-based
and nine transformer-based acoustic foundation models
against a convolutional baseline model,
testing their performance on arousal, valence, dominance, and emotional category classification.
The test results highlight,
that models with high correlation or recall
might rely on shortcuts
-- such as text sentiment --,
and differ in terms of fairness.
\end{abstract}

\glsresetall

%%%%%%%%%%%%%%%%%%%%%%%%%%%%%%%%%%%%%%%%%%%%%%%%%%%%%%%%%%%%%%%%%%%%%%%%%%%
\section{Introduction}
\label{sec:introduction}

Machine learning models are developed to fulfill a given objective
by presenting them examples.
In \ac{SER}
we might have the prediction of arousal as objective
and audio samples
with an associated arousal value
as examples.
Development pipelines are restricted
to certain model specifications
and a limited amount of examples
leading to non-perfect fulfillment of the objectives.
This requires further evaluation steps
to estimate the performance of the models.
The evaluation can focus on tracking progress
in a given field by specifying benchmarks
in order to compare models~\citep{thiyagalingam2022scientific}.
%As much of the progress in recent times
%is achieved by general-purpose foundation models~\citep{bommasani2021opportunities}
%the focus has shifted towards more and smaller datasets
%as benchmarks \citep{wang2018glue}
%and a larger variety of tasks \citep{turian2022}.
However, two models,
showing the same accuracy performance,
might have different general behaviour and properties,
due to the underspecification of the applied development pipeline~\citep{damour2022underspecification},
which can lead to models
containing spurious correlations
or learnt shortcuts.
For applications of models,
therefore,
it is important
to understand their behaviour, robustness, and potential biases~\citep[][§4.4]{bommasani2021opportunities},
and communicate those to stakeholders~\citep{mitchell2019modelcard}.

%This means two models showing the same accuracy performance in a benchmark
%might have very different general behaviour and properties,
%that need to be understood~\citep[][§4.4]{bommasani2021opportunities}
%and communicated to stakeholders~\citep{mitchell2019modelcard}.

As it is most often required that a model stays within a certain range
of expected behaviour,
testing is a valid evaluation approach, as it can detect
differences between existing and required behaviour~\citep{zhang2020}.
Testing machine learning models provides a greater challenge
compared to software testing~\citep{ammann2016}
as the models provide answers to questions
for which no previous answer -- e.\,g., label --
was available~\citep{murphy2007testing}.
This can be solved by changing available input samples from test sets
in a way that labels are preserved~\cite{tian2018deeptest}
or with an expected change of labels~\citep{ribeiro2020checklist}.
In addition, synthetic data with known labels might be generated~\citep{zhou2020deepbillboard}. 

% Test coverage
%
% We have to decide if we want to include this.
% At the moment I think the introduction is already long enough.
%
% The effect of most properties of machine learning models
% is only visible when testing the model as a whole system,
% which makes it very hard to apply the concept of unit tests.
% If access to the neurons of the model is available,
% which is for example not the case
% if the model comes in the form
% of an ONNX file\footnote{\url{https://onnxruntime.ai/}},
% tests can try to apply changes to input signals
% until a full or very high neuron coverage is reached~\citep{pei2019}.

In \ac{SER} no general testing approach was proposed so far.
The Computational Paralinguistics Challenges~\citep{schuller2019compare}
tracked progress for \ac{SER}
and have led to rapid progress in the area.
\citet{scheidwasser-clow2022} have introduced a multi-dataset benchmark
focusing on the evaluation of fine-tuned foundation models. 
%BS: added: 
Further, 
\citet{jaiswal2021robustness} have evaluated the robustness of a model
that predicts categorical emotions under different data augmentations.
Their model under test showed significant performance degradation
for most of the applied augmentations.
They also showed that some of the augmentations like change in pitch,
adding laughter, crying, or speeding up the utterance
can affect the underlying label as well.
\citet{triantafyllopoulos2022} presented a framework to estimate
how much the prediction of valence of a \ac{SER} model
depends on the extracted sentiment from text
instead on the tone of voice.
\citet{schmitz2022bias} addressed the topic of fairness~\citep{bellamy2018ai},
which is otherwise rarely addressed in the \ac{SER} community so far.

%BS: added some commas throughout ;)
In this paper, we follow \citet{zhang2020} and propose to implement
the evaluation of model behaviour in the form of offline tests.
The development of the tests is not driven by the concept of unit tests
or coverage, but motivated by possible applications of the model,
and how it should behave under certain conditions.
We focus on \ac{SER} models
with the regression task of predicting emotional dimensions
(arousal, dominance, valence)
and the classification task of predicting emotional categories
(anger, happiness, neutral, sadness).
The tests focus on acoustic foundation models
including
HuBERT~\citep{hsu2021hubert},
wav2vec 2.0~\citep{baevski2020wav2vec},
wavLM~\citep{chen2022wavlm},
data2vec~\citep{baevski2022data2vec},
and xLSTM~\citep{beck2024xlstm}
architectures.
We test the model behaviour in terms of model correctness,
fairness, and robustness.

% Correctness involves tests requiring a given accuracy for tasks
% like ranking of speakers, 
% or matching a ground truth distribution.
% Fairness tests require the models to show no biases
% regarding protected groups,
% for example based on accent or sex.
% Robustness tests require model results to be independent
% of changes in the recording conditions,
% like different microphones or additive noise.
For fairness tests we show how to select the most critical test thresholds
possible,
given the test datasets,
and discuss for other tests
how to select thresholds based on the desired model application.
The implementations and detailed results of all tests are available at \url{https://audeering.github.io/ser-tests/}.

%We solve the test oracle problem
%for test data without labels
%by comparing predictions across protected groups,
%e.\,g., assuming that we should get the same distribution of model predictions
%for groups that differ by a certain aspect
%like speaker accent;
%or by generating different versions of labelled data
%with data augmentations
%for which we know they should not affect the label.

%%%%%%%%%%%%%%%%%%%%%%%%%%%%%%%%%%%%%%%%%%%%%%%%%%%%%%%%%%%%%%%%%%%%%%%%%%%
\section{Method}
\label{sec:method}

\citet{zhang2020} group testing properties for machine learning systems into
the categories correctness, model relevance, robustness, security, data privacy, efficiency, fairness, and interpretability.
In this work, we focus on the properties
\textbf{correctness},
\textbf{robustness},
and \textbf{fairness}.
Correctness tests measure the extent to which
the model predicts the correct label under test,
robustness tests investigate the model behaviour
in the presence of perturbations, and
fairness tests check
whether the model has biases against certain attributes.
We do not cover model relevance,
i.\,e., whether the complexity of the model fits the data,
interpretability,
and data privacy, as they are most commonly covered
by means of white-box and grey-box testing
and may be architecture specific,
whereas we require our framework
to include only universally applicable black-box tests.
Security testing is often related to adversarial robustness,
especially for tasks such as autonomous driving,
where facing adversarial examples
introduces high risks~\citep{zhou2020deepbillboard}.
For the task of speech emotion recognition,
we assume that the model is only applied in scenarios
with no or low risk and do not cover security in addition to robustness.
We do not address efficiency with a dedicated set of tests,
but the results of all tests can be used
to compare smaller variants of a model~\citep{ren2022fast}
or models trained on a subset of the training data~\citep{spieker2019towards}
of its original version.

For each test
we propose a selection of evaluation metrics,
and suggest a threshold that determines a passing or failing result.
We propose a method to set the fairness test thresholds
automatically based on numeric simulations
and independent of the application.
% AD: Changed to address review comment "Lines 20-22 in Page 2 should be rewritten"
The suggested thresholds for correctness and robustness
should be adjusted depending on the requirements of an application.
For example, one could aim for a \ac{CCC} of $0.6$
for all included test sets before deploying a model
and use the percentage of passed tests
to track progress towards that goal.
If no such requirement is defined,
we recommend to only focus on the individual test results
without enforcing thresholds instead.
% Whereas the thresholds for correctness
% and robustness need to be defined
% with certain applications in mind.
In general,
we suggest to use these tests
as tools to understand model behaviour
instead of only looking at the final test scores,
since in some cases
a small change in a metric or threshold
could be the difference between a passing and failing test.

%==========================================================================
\subsection{Datasets}

For our tests, we include a multitude of emotional datasets from various domains.
In order to compare results between datasets,
we map the categorical labels to a standardised set of names
(e.\,g., anger, happiness, neutral, or sadness).
We map samples labelled as joy to the category of happiness.
On the dimensional labels
we apply min-max scaling to scale them to the range of $[0,1]$.
If a dataset has enough speakers and data points,
but does not contain a test split,
we define one.
This allows to include corresponding datasets in training,
if desired.

The \acf{crema-d}~\citep{cao2014crema} is a dataset
containing 7,442 samples
from 91 actors (48 female, 43 male)
between the ages of 20 and 74\,years
from various races and ethnicities.
The actors were asked to portray a selection of 12 English sentences
in different emotions (anger, disgust, fear, happiness, neutral, and sadness)
in varying levels of intensity.
Emotion ratings are available for the audio modality,
the visual modality,
and for the combined modalities.
We use the ratings from the audio modality,
removing samples with no agreement from the dataset.
As there is no official test set,
we define our own split of the dataset with the samples from 17 speakers
(8 female, 9 male).
The speakers were first grouped
based on their gender, race, and ethnicity,
and then each group was randomly split.
~\citet{wierstorf2023} document the selected files. %BS: this should be made reproducible, e.g., by providing the speaker IDs in a footnote or a download link for paritions, such as a github repository or alike...
% We could add the specific speakers we use for our test splits for reference,
% but this would take up some space.
% Maybe could be added in an appendix?
% HW: published splits on zenodo and added them here as reference.

\acf{des}~\citep{engberg1997} is an approximately 30 minutes long dataset
containing recordings in Danish
from 4 actors (2 female, 2 male)
who convey the emotions anger, happiness, neutral, sadness, and surprise.
We use the entire dataset as the test set.

\acf{emodb}~\citep{burkhardt2005emodb} is a German dataset
in which 10 actors (5 female, 5 male)
each portray 10 sentences
in the emotions anger, boredom, disgust, fear, happiness, neutral, and sadness. Recordings took place in an anechoic chamber.
We select randomly 2 female and 2 male speakers for our test split,
with the constraint of having a similar distribution of emotional classes
between the splits~\citep{wierstorf2023}.
%BS: again - this is non-reproducible and arbitrary - can you please provide reaosning (random is fine) for choices here and potnetially below and access to the partitionings.

\acf{emovo}~\citep{costantini2014emovo} is an Italian dataset
with recordings of 6 actors (3 female, 3 male)
tasked to portray 14 sentences
in the emotional states anger, disgust, fear, joy, neutral, sadness, and surprise.
We use the entire dataset for our tests.

\acf{iemocap}~\citep{busso2008iemocap} is a corpus
collected by the Speech Analysis and Interpretation Laboratory (SAIL)
at the University of Southern California (USC).
10 actors were recorded in dyadic sessions
which include a scripted portion
and an improvised portion,
designed to elicit emotional data,
resulting in approximately 12 hours of speech.
Each of the 5 recorded sessions
contains a male and a female speaker.
The dataset has been annotated for the arousal, dominance, and valence dimensions
as well as for categories
(anger, disgust, excitement, fear, frustration, happiness, neutral, other, and sadness).
We form a test split of the dataset from sessions 4 and 5,
but use all sessions for certain tests
where we require a higher number of speakers to gain relevant results~\citep{wierstorf2023}.

\acf{meld}~\citep{poria2019meld} is an enhancement
on the EmotionLines dataset~\citep{hsu2018emotionlines}
that extends it to include audio and visual modalities.
It contains about 13,000 utterances
from the English TV-series \textit{Friends},
and has been annotated with emotion and sentiment labels.
The emotion categories are anger, disgust, fear, joy, neutral, sadness, and surprise.
We use the official test set,
but removed files shorter than $0.76$\,s
or longer than $30$\,s~\citep{wierstorf2023}.

\acs{msppodcast}~\citep{lotfian2019msppodcast} is a large speech emotional dataset
built from segments from English podcast recordings.
The dataset is annotated using crowdsourcing
for the arousal, dominance, and valence dimensions,
and categorical labels
(anger, contempt, disgust, fear, happiness, neutral, other, sadness, and surprise).
We use version 1.7 of the dataset,
which has roughly 100\,hours of speech data.
We evaluate our tests on both of the test sets:
test set 1 with 30 male and 30 female speakers
and test set 2 with approximately 3,500 segments
from 100 podcasts not used in other partitions.

\acf{pesd}~\citep{pesd2020} comprises 240 recordings in Polish
from 8 actors (4 female, 4 male).
Each speaker utters 5 sentences with 6 types of emotional prompts:
anger, boredom, fear, joy, neutral, and sadness.
We use the combined set of samples as a test set.

\acf{ravdess}~\citep{livingstone2018ryerson} contains audio and video recordings
from 24 professional actors (12 female, 12 male)
vocalising 2 English statements in speech and song.
Speech samples are expressed
in the emotions anger, calm, disgust, fear, happiness, neutral, sadness, and surprise. 
Each emotion except for the neutral expression
is produced in a normal as well as a strong intensity.
For our test set, we select randomly 2 female and 2 male speakers
and only use their speech samples~\citep{wierstorf2023}.

Some tests might use additional datasets,
e.\,g., to add background noise.
In those cases,
the dataset is introduced in the test definition.

%==========================================================================
\subsection{Correctness Tests}

%AD: adjusted table by removing metrics:
% Mean Value, MDE, Class Prop. MDE, Precision Top/Bottom, Top-Bottom Confusion
\begin{table*}[t]
	\centering
	\caption{Overview of the correctness tests, their test sets, metrics, and passing conditions.}
	\label{tab:correctness_thresholds}
\begin{tabular}{p{0.2\textwidth}p{0.075\textwidth}p{0.3\textwidth}p{0.2\textwidth}p{0.075\textwidth}}
\toprule
\textbf{Test} & \textbf{Task} & \textbf{Test Sets} & \textbf{Metric} & \textbf{Condition}                           \\
\midrule
Correctness Classification & categories & \acs{crema-d}, \acs{des}, \acs{emodb}, \acs{emovo}, & \ac{PPC} &                     $>0.5$ \\
                            &            & \acs{iemocap}, \acs{meld}, \acs{msppodcast} (test-1), & \ac{RPC} &                     $>0.5$ \\
                            &            & \acs{msppodcast} (test-2), \acs{pesd}, \acs{ravdess} & \ac{UAP} &                     $>0.5$ \\
                            &            &  & \ac{UAR} &                     $>0.5$ \\
Correctness Consistency & dimensions & \acs{crema-d}, \acs{des}, \acs{emodb}, \acs{emovo}, \acs{iemocap}, \acs{meld}, \acs{pesd}, \acs{ravdess} & Samples in Expected Range &                    $>0.75$ \\
Correctness Distribution & categories & \acs{crema-d}, \acs{des}, \acs{emodb}, \acs{emovo}, \acs{iemocap}, \acs{meld}, \acs{msppodcast} (test-1), \acs{msppodcast} (test-2), \acs{pesd}, \acs{ravdess} & Relative Diff. Per Class &  $\lvert\cdot\rvert <0.15$ \\
                            & dimensions & \acs{iemocap}, \acs{msppodcast} (test-1), &  Jensen-Shannon Distance &                     $<0.2$ \\
                            &            & \acs{msppodcast} (test-2) & \\
Correctness Regression & dimensions & \acs{iemocap}, \acs{msppodcast} (test-1), \acs{msppodcast} (test-2)& \ac{CCC} &                     $>0.5$ \\
                            &            &   & \ac{MAE} &                   $<0.1$ \\
                            &            &                                                      & \ac{PCC} &                     $>0.5$ \\
Correctness Speaker Average & categories & \acs{iemocap}, \acs{meld}, \acs{msppodcast} (test-1) & Class Proportion MAE &                     $<0.1$ \\
                            & dimensions & \acs{iemocap} (full), \acs{msppodcast} (test-1), & MAE &                     $<0.1$ \\
Correctness Speaker Ranking & categories & \acs{meld}, \acs{msppodcast} (test-1) & Spearmans Rho &   $\lvert\cdot\rvert >0.7$ \\
                            & dimensions & \acs{msppodcast} (test-1), \acs{msppodcast} (test-2) & Spearmans Rho &   $\lvert\cdot\rvert >0.7$ \\
\bottomrule
\end{tabular}
\end{table*}

The correctness tests require
that the model predictions follow the true labels as closely as possible.
Correctness is a fundamental goal of most machine learning systems
and as such rarely overlooked.
However,
special care is required to distinguish
between different types of errors,
some of which may matter more to a user than others.
By looking at correctness from different viewpoints
and with different metrics,
more nuanced insights in model behaviour can be gained.

\cref{tab:correctness_thresholds} shows a summary of the discussed correctness tests.
Furthermore, it lists the test sets we apply the test metrics on,
as well as the passing conditions we apply for each test metric.
The thresholds for the passing conditions provide only an example
to present our testing framework
and to show how the test results
can be summarised into a percentage of passed tests.
We suggest to adjust the thresholds
to the needs of each individual application,
or to compare the metric results directly
without enforcing a binary result of passing or failing.
An alternative option would be
to base the thresholds for correctness on average human rater performance,
e.\,g. by randomly selecting a rater for each sample,
or by averaging all individual raters' performance.
%AD adjusted number to removed metrics
% The number of correctness tests are 66 for arousal,
% 72 for dominance,
% 76 for valence,
% and 196 for categorical emotions.
The number of correctness tests are 54 for arousal,
60 for dominance,
64 for valence,
and 160 for categorical emotions.
In the following,
we discuss correctness tests
that need additional information to what is presented in  \cref{tab:correctness_thresholds}.

% --- Correctness Classification and Regression
The \textbf{Correctness Classification} tests
and \textbf{Correctness Regression} tests
include standard metrics of correctness.
We include \acf{UAR}, \acf{UAP}, \acf{RPC}, and \acf{PPC} to evaluate classification models
and 
\acf{CCC}, \acf{PCC}, and \acf{MAE} to evaluate
regression models.

% --- Correctness Consistency
The \textbf{Correctness Consistency} tests 
check whether the models’ predictions
on dimensional tasks
are consistent with the expected result
for samples with certain categorical labels.
For example,
happiness is characterised by high valence
and fear tends to coincide with low dominance.
Based on comparing various literature
results~\citep{fontaine2007, gillioz2016mapping, hoffmann2012mapping, verma2017affect},
we expect a correspondence between dimensional values
and emotional categories as presented in \cref{tab:correspondence},
% and \cref{fig:correctness-consistency},
where dimensional values $\ge 0.55$ are counted to be in the high range, %BS: these ranges seem arbitrary - some reasoning or discussion on that automatically tuning these could lead to different outcomes, perhaps?
%BS: also, visualising this area (3D-plot?) could add to the visual appearance - as is, it's mostly text :)
values between $0.3$ and $0.6$ in the neutral range,
and values $\le 0.45$ in the low range.
% AD: changed to address "mentions about Fig. 5. It seems out of context." --> decided to remove reference to Fig. 5 completely
% The expected ranges are also visualized
% for a model's results in \cref{fig:correctness-consistency}.
% AD: added to address "Reviewer 2: ranges for correctness consistency test overlap"
Note that the values we referenced
in literature varied in range,
and sometimes no clear correspondence was found.
We use overlapping ranges
to allow for some variance in results,
and only penalize predictions
that clearly lie outside the expected range.
% We then evaluate dimensional models for predictions
% on categorical datasets
% by checking the percentage of the predictions
% within the expected low, neutral, and high ranges.
% We include all categorical datasets
% except for \ac{msppodcast} for this test,
% as we have found some inconsistencies
% between the ground-truth dimensional
% and the ground-truth categorical labels on this dataset
% (for instance we observed a large portion of the samples
% labelled as fear with high dominance values).

\begin{table}[t]
    \centering
    \caption{
    Correspondence between emotional categories and dimensional values
    based on literature review.
    }
    \begin{tabular}{llll}
        \hline
        \textbf{emotion} & \textbf{valence} & \textbf{arousal} & \textbf{dominance} \\
        \hline
        anger       & low     & high    & high    \\
        boredom     & neutral & low     &         \\
        disgust     & low     &         &         \\
        fear        & low     & high    & low     \\
        frustration & low     &         &         \\
        happiness   & high    &         & neutral \\
        neutral     & neutral & neutral & neutral \\
        sadness     & low     & low     & low     \\
        surprise    &         & high    & neutral \\
        \hline
    \end{tabular}
    \label{tab:correspondence}
\end{table}

% --- Correctness Distribution
% When calculating the Jensen-Shannon divergence~\citep{endres2003}
% for the \textbf{Correctness Distribution} test,
% we bin the distributions into 10 bins.

The \textbf{Correctness Distribution} tests ensure
that the distributions of the model predictions
are very similar to the gold standard distributions.
The Jensen-Shannon Distance,
i.\,e., the square root of the Jensen-Shannon divergence
measures the similarity between two random distributions~\citep{endres2003}.
Values range from $0$ to $1$,
with lower values indicating more similar distributions.
We bin the distributions into 10 bins and then calculate the distance.
In addition, we calculate
the absolute difference between the two distributions' mean value.
For categorical models,
we test whether the number of samples per class is comparable
between the model prediction and the gold standard.
We measure the difference of the number of samples
in relative terms compared to the overall number of samples,
for each class.

% --- Correctness Regression
% The \textbf{Correctness Regression} tests
% include the standard metrics used to evaluate regression problems:
% \ac{CCC}, \ac{PCC}, and \ac{MAE}.
% We did not use \ac{MSE} or \ac{RMSE}, but these
% metrics could be added
% to the test for applications where large errors
% should be given a higher weight. 
% %BS: Why not MAE (!!!)? This is imho the most intuitive and potentially most used one! Also RMSE?
% %AD changed from MSE to MAE

% --- Correctness Speaker Average
Certain applications of \ac{SER} models may be interested
in the average emotional value for each speaker.
The \textbf{Correctness Speaker Average} tests
check whether the models' estimate of the average speaker value
is close to the truth.
For regression,
we measure the \acf{MAE}
%AD: removed metric
% and \acf{MDE}
%BS: ah, voilà - MAE ;) Please introduce and reuse proerly as abbreviation and reuse below.
between the true speaker average
% AD: changed to address "the authors mention about "predicted speaker average". How is this defined and how is it computed?"
and the predicted speaker average,
i.\,e., the average of all predictions
of samples belonging to a speaker.
For classification,
as we cannot compute a single average value,
we compare the proportions of samples
that are assigned a certain class per speaker.
We then calculate the \ac{MAE}
%AD: removed metric
% and the \ac{MDE}
in the estimated proportion of samples
for each class.
We only consider speakers with at least 10 samples for regression,
and with at least 8 samples per class for classification.
We apply the tests on test sets for which six or more speakers remain
that fit the criteria.
% For dimensions, we use all available dimensional datasets, 
% but take the entire \ac{iemocap} dataset %BS: added - please check:
% as test, 
% since we require a higher number of speakers for this test.
% For categorical emotion,
% we also include the full \ac{iemocap} set,
%BS: added - please check:
% as test 
% as well as the test set of \ac{meld}
% and the test set 1 of \ac{msppodcast}.
%BS: can this be perhaps best be put into a table (what test sets for what)? Perhaps also a small table on measures used for classification/regression - readers love tables and figures :)

% --- Correctness Speaker Ranking
We test a potential ranking of speakers
based on their average,
for instance to spot outliers on either side of the ranking.
This is covered in the \textbf{Correctness Speaker Ranking} tests.
The average values
(as computed in the Correctness Speaker Average tests)
are used to create a ranking.
For classification, we create a separate ranking for each class label,
e.\,g., for the anger class, we rank speakers
with a higher proportion of anger samples higher,
and speakers with a lower proportion of anger samples lower.
%AD: adjusted to better explain this metric
As a measure of the correctness of the ranking, 
we use \ac{spearmans}.
% As a measure of the overall ranking, 
% we use \ac{spearmans}.

%AD: removed metric
% The top and bottom of a ranking are often of particular interest,
% which we test by computing the precision
% of the upper and lower quantiles of the ranking
% (bottom $25\%$ and top $25\%$).
% Finally,
% we consider Top Bottom Confusions,
% i.\,e., speakers that are ranked into the top (or bottom) quantile
% although they are in the opposite quantile
% in the true ranking.
% The tests are applied on the two \ac{msppodcast} test sets
% for dimensions
% and on \ac{msppodcast} test set 1
% and \ac{meld} for categories
% (leaving out \ac{iemocap} due to speaker averages that lie very close together).

%==========================================================================
\subsection{Fairness Tests}

Despite a long history of research in the field,
a universal definition for fairness has not been established,
neither in a general sense,
nor when applied to machine learning~\citep{mehrabi2021survey}.
Many widely used definitions of fairness state
that no bias should exist
for certain protected attributes~\citep{corbett2018measure}.
In \citet{mehrabi2021survey}, algorithmic fairness
is grouped into three main types:
\textit{individual fairness},
which aims to give similar predictions to similar individuals,
\textit{group fairness},
which tries to treat different protected groups equally,
and \textit{subgroup fairness},
which combines the two previous approaches
by selecting a group fairness constraint
and checks whether the constraint applies across sets
of combinations of protected attribute values. 
Individual fairness would require data with similar samples
that differ only in the protected attribute,
and subgroup fairness would require test sets
with annotations for all types of fairness groups
(e.\,g., accent or language) at the same time,
both of which are not easily available.
Thus,
we employ different types of group fairness tests
and distinguish between cases
where the ground truth emotion label is known
and where it is not. 
Note that the fairness tests should not be interpreted as proof
that a model is fair
when it passes the tests,
but rather as an indicator that the model
is likely not fair towards a certain protected group
when it fails the tests.
The tests in the following do not cover all relevant groups
for which fairness should be considered,
but they provide a start to be expanded on. 

\emph{Statistical parity}
(or \emph{demographic parity})
is a group fairness criterion
which enforces
that the model function \(f(X,S)\),
given the input data \(X\)
and the protected group \(S\),
is statistically independent of \(S\)~\citep{del2020review}.
For the classification of classes \(c \in C\),
this is given when for all \(s \in S\)

\begin{equation}
    \mathbb{P}(f(X,S)=c) = \mathbb{P}(f(X,S)=c | S=s),
\end{equation}
% AD: added to address "The authors should explain P in Eq. (1) - although obvious."
where $\mathbb{P}(\cdot)$ denotes the probability
of an event.

We apply this to our tests
for unlabelled data
by comparing the class wise distributions of samples
for the different group members,
and require that the differences
are below a given threshold.
We refer to this metric as the relative difference per class.
For a regression model with \(f(X,S) \in [0,1]\),
statistical independence requires
that for all \( s \in S\) and \(z \in [0,1]\)

\begin{equation}
    \mathbb{P}(f(X,S) \geq z)
    = \mathbb{P}(f(X,S) \geq z | S=s).
\end{equation}

We follow \citet{agarwal2019fair}
and discretise this requirement
by binning the model outputs into evenly spaced bins \(\mathcal{Z}\).
We then require for the binned model output \(f_{\text{bin}}(X,S)\)
and for all \(s \in S\) and \(\bar{z} \in \mathcal{Z}\)

\begin{equation}
    \mathbb{P}(f_{\text{bin}}(X,s) \geq \bar{z}) =
    \mathbb{P}(f_{\text{bin}}(X,S) \geq \bar{z} | S=s).
\end{equation}

In order to get a clearer insight
into which regions of the output space contain disparities,
we reformulate the requirement
to check the probability of the intervals corresponding to each individual bin.
We refer to this metric as the relative difference per bin.
In our tests, we use four bins.
For statistical parity
the distribution of the prediction
should be independent of the protected group.
That is why we also check
whether the shift in mean value
is below a threshold as an indicator of fairness.

If a dataset has ground truth labels for emotion,
we test the correctness for each group member
(similar to the bounded group loss in \citet{agarwal2019fair})
and require that the difference in overall performance
is low in terms of \acf{CCC}
% AD: removed metric
% and \ac{MAE}
for regression
and in terms of \acf{UAR}
% AD: removed metric
% and \acf{UAP}
for classification.

The criterion of \emph{Equalised Odds}~\citep{mehrabi2021survey} states
that all members of a group
should have equal rates for true positives and false positives
in a binary classification scenario.
We apply this to multi-class classification
by comparing the difference in \acf{RPC} and \acf{PPC}.
For regression, we again map the model outputs into four evenly spaced bins
and compare the difference in recall per bin and precision per bin.
% AD: removed metric
% We also consider the \ac{MDE} for regression
% in order to see if there is a bias in the mean
% towards a certain direction for a protected group.

\begin{figure}
    \centering
    \includegraphics[width=\columnwidth]{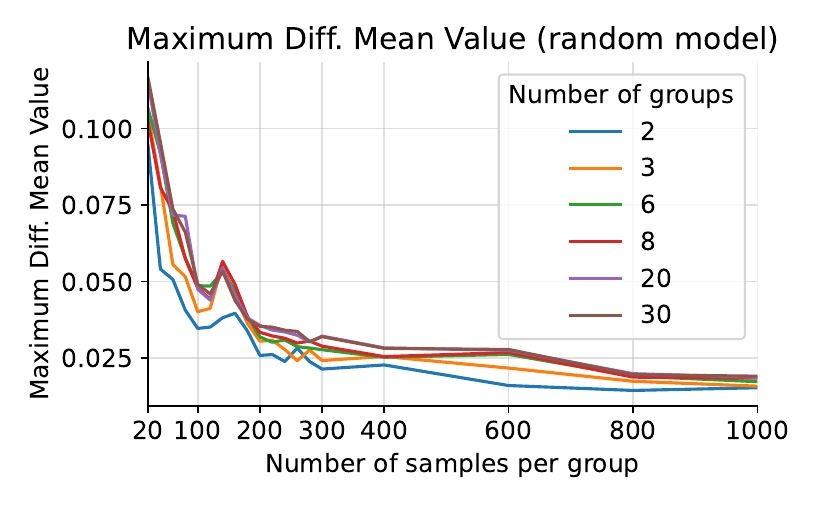}
    \caption{
        Maximum difference in mean value
        among 1000 simulations
        with a varying number of groups
        and number of samples per group.
    }
    \label{fig:simulate_mean_value}
\end{figure}

For all fairness tests,
% AD: change wording to clarify for Reviewer 1: "what are these random models?"
we simulate models with random behaviour
by generating random values
and use their test results as reference
for setting test thresholds.
The motivation behind this approach is
that a model that just samples values from
a distribution has no bias towards certain groups.
% we use simulations based on random models as reference
% for setting test thresholds.
% A model that outputs purely random predictions
% has no bias towards certain groups.
However, when the number of samples in the test set is small,
the change in prediction for a protected group
has a higher chance of being high, even for a random model.
Thus,
we simulate potential test outcomes for random models
under different conditions.
For regression tasks,
we generate random model samples from a truncated Gaussian distribution
with values between $0$ and $1$,
a mean value of $0.5$,
and a standard deviation of $\frac{1}{6}$.
For categories, 
our random model samples from a uniform categorical distribution
with four categories.
For test metrics
where the ground truth values are taken into consideration,
we also generate the ground truth values randomly
%BS: "randomly" - reproducible? provision of random seed and function, perhaps?
by sampling values from a truncated Gaussian distribution
for continuous values.
For categories,
we consider both a uniformly distributed ground truth,
as well as a sparsely distributed ground truth
(with class probabilities $(0.05, 0.05, 0.3, 0.6)$).
Each simulation is repeated 1000 times
and we use the maximum difference in prediction
for a protected group as a reference value for our test threshold.
The simulation results are shown
% AD: changed to difference in mean value since metric diff mae was removed
for the difference in mean value
% for the example of the difference in mean absolute error 
%BS: added:
as 
test metric
in Fig.~\ref{fig:simulate_mean_value}.
% AD: added to clarify our approach for Reviewer 1
For example, if we have a test dataset with
3 protected groups and at least 600 samples per group,
we can see that the maximum difference in mean value
is below $0.025$,
and we can use this value as a threshold
when applying the test
on this dataset.
% ----
For certain test sets,
we encountered the issue that the distribution
of the ground truth for certain groups
varies considerably from the distribution of other groups.
The maximum difference in prediction in the simulation increases
when the ground truth labels show a bias for a particular group.
In order to avoid this,
we balance the test sets by selecting 1000 samples from the group
with the fewest samples,
and 1000 samples from each other group with similar truth values.
For regression tasks,
this may result in certain bins having very few samples.
In these cases,
we decided to skip bins with too few samples.
Specifically,
we set the minimum number of samples $n_\text{bin}$
to the expected number of samples in the first bin
for a Gaussian distribution with a mean of $0.5$
and a standard deviation of $\frac{1}{6}$:

\begin{equation}
    n_{\text{bin}} = \mathbb{P}(X\leq0.25) \cdot n,
    \label{eq:min_samples_per_bin}
\end{equation}

\noindent
where $n$ is the total number of samples,
and the random variable $X$ follows the aforementioned distribution.
We take the same approach for the tests
with unlabelled test sets in the case
that a model has very few predictions in a certain bin for the combined test set. With this, we avoid that a change in prediction of only one sample
in a certain bin for one group
could result in a large difference in the test metric.
% AD: added to clarify for Reviewer 1
% how the approach can be applied to other tasks
This approach of setting thresholds
for fairness test metrics
can also be applied to other tasks by
adjusting the distribution
the random predictions are sampled from
(e.\,g., using a uniform categorical distribution for 5 classes instead of 4 classes).

%AD: adjusted table by removing metrics:
% Diff MAE, Diff MDE, Diff UAP
\begin{table*}[t]
	\centering
	\caption{Overview of the fairness tests, their test sets, metrics, and passing conditions. For metrics involving bins, the minimum number of samples per bin $n_\textnormal{bin}$ is shown in parentheses. Bins with fewer than $n_\textnormal{bin}$ are skipped in the test.}
	\label{tab:fairness_thresholds}
\begin{tabular}{p{0.2\textwidth}p{0.075\textwidth}p{0.25\textwidth}p{0.25\textwidth}p{0.075\textwidth}}
\toprule
\textbf{Test} & \textbf{Task} & \textbf{Test Sets} & \textbf{Metric} & \textbf{Condition}                            \\
\midrule
Fairness Accent & categories & speech accent archive & Relative Diff. Per Class &  $\lvert\cdot\rvert <0.225$ \\
             & dimensions & speech accent archive & Diff. Mean Value &  $\lvert\cdot\rvert <0.075$ \\
             &            &                                                 & Relative Diff. Per Bin ($n_\text{bin}=4$) &  $\lvert\cdot\rvert <0.225$ \\
Fairness Language & categories & Mozilla Common Voice & Relative Diff. Per Class &    $\lvert\cdot\rvert <0.1$ \\
             & dimensions & Mozilla Common Voice & Diff. Mean Value &   $\lvert\cdot\rvert <0.03$ \\
             &            &                                                 & Relative Diff. Per Bin ($n_\text{bin}=67$) &    $\lvert\cdot\rvert <0.1$ \\
Fairness Linguistic Sentiment & categories & \textsc{CheckList} (synthesized) & Diff. Class Proportion Shift &  $\lvert\cdot\rvert <0.075$ \\
             & dimensions & \textsc{CheckList} (synthesized) & Diff. Bin Proportion Shift ($n_\text{bin}=67$) &  $\lvert\cdot\rvert <0.075$ \\
             &            &                                                 & Diff. Mean Shift &  $\lvert\cdot\rvert <0.025$ \\
Fairness Pitch & categories & \acs{msppodcast} (test-1) & Diff. PPC &    $\lvert\cdot\rvert <0.1$ \\
             &            &                                                 & Diff. RPC &  $\lvert\cdot\rvert <0.225$ \\
             &            &                                                 & Diff. UAR &  $\lvert\cdot\rvert <0.075$ \\
             & dimensions & \acs{msppodcast} (test-1) & Diff. CCC &    $\lvert\cdot\rvert <0.1$ \\
             &            &                                                 & Diff. Precision Per Bin ($n_\text{bin}=67$) &  $\lvert\cdot\rvert <0.125$ \\
             &            &                                                 & Diff. Recall Per Bin ($n_\text{bin}=67$) &  $\lvert\cdot\rvert <0.125$ \\
Fairness Sex & categories & \acs{iemocap}, \acs{msppodcast} (test-1) & Diff. PPC &  $\lvert\cdot\rvert <0.075$ \\
             &            &                                                 & Diff. RPC &  $\lvert\cdot\rvert <0.175$ \\
             &            &                                                 & Diff. UAR &  $\lvert\cdot\rvert <0.075$ \\
             & dimensions & \acs{iemocap} (full), \acs{msppodcast} (test-1) & Diff. CCC &  $\lvert\cdot\rvert <0.075$ \\
             &            &                                                 & Diff. Precision Per Bin ($n_\text{bin}=67$) &    $\lvert\cdot\rvert <0.1$ \\
             &            &                                                 & Diff. Recall Per Bin ($n_\text{bin}=67$) &    $\lvert\cdot\rvert <0.1$ \\
\bottomrule
\end{tabular}
\end{table*}

\cref{tab:fairness_thresholds} gives an overview of the discussed fairness
tests, and also lists the applied test sets and passing conditions.
All fairness thresholds are based on the previously described
simulations with random models
and depend on the number of protected groups in the used test,
the number of samples per group, as well as whether the test data is
distributed sparsely in case of categories.
For some test sets certain regression
bins may have fewer samples than expected for our assumption of a
Gaussian distribution with a mean of $0.5$
and a standard deviation of $\frac{1}{6}$ of the ground truth in labelled sets, and of the prediction in unlabelled sets. 
Therefore, we also list the minimum number of samples per bin $n_\text{bin}$
(see Eq.~\ref{eq:min_samples_per_bin}) in parentheses behind the respective metrics.
% Could add the following if needed for clarity?
% For instance, for the Fairness Pitch test, we compute the expected minimum
% number of samples in a bin by inserting the number of samples per group, 1000, in Eq.~\ref{eq:min_samples_per_bin}:
% \begin{equation*}
%     n_{\text{min}} = \mathbb{P}(X\leq0.25) \cdot 1000 \approx 67,
% \end{equation*}
% and we skip bins in the test with fewer than 67 samples.
When possible, test sets with a large number of samples per protected
group should be used, as the tests can lead to more meaningful
conclusions.
For instance, for the Fairness Accent tests, the thresholds are based on
simulations with a random uniform categorical and random Gaussian model for 30 fairness groups
and only 60 samples per group, and are thus relatively high.
For all other tests, we can base our thresholds on simulations with at least 1000
samples per group. Note that for the Fairness Language tests, we increased the calculated thresholds slightly to accommodate for potential variations
of the content
and recording context across different languages in the database.
% AD: Adjusted numbers for removed metrics
% The number of fairness tests are 557
% for each arousal, dominance, valence,
% and 520 for categorical emotions.
The number of fairness tests are 348
for arousal, 352 for dominance,
344 for valence,
and 307 for categorical emotions.
In the following paragraphs,
we provide additional information to the single tests,
besides what is shown in \cref{tab:fairness_thresholds}.

% --- Fairness Accent
The \textbf{Fairness Accent} tests use randomly selected samples
from the speech accent archive~\citep{weinberger2015},
which contains more than 2000 speech samples
from speakers of different nationalities and native languages.
All speakers read the same paragraph in English,
lasting a little over 3 minutes per recording for most cases.
We use the recordings from 5 female and 5 male speakers for each accent
(including native English).
We compare the predictions of each of 31 accents
to the predictions of the combined data.
% % We base the thresholds on simulations with a random categorical
% % and random Gaussian model for 30 fairness groups and 60 samples per group. %BS: "random" -- reproducible? Please make it reproducible if possible (random seed, repository, ...)
% We measure the difference in mean value and the relative
% difference per bin 
% for regression tasks,
% and the relative difference per class for classification. %BS: these feel very cumbersome - where do these thresholds come from?
% % For the test metric relative difference per bin
% % we require at least 4 predictions per bin in the combined test set,
% % or we skip that bin.

% --- Fairness Language
In the \textbf{Fairness Language} tests we use 2000 randomly selected samples
from Mozilla Common Voice~\citep{commonvoice2020}
for each of the languages German,
English,
Spanish,
French,
Italian,
and Mandarin Chinese. %BS: you really need to do a repository, please - way too much random everywhere...
The prediction of each individual language
is compared to the combined data.
% % The thresholds for this test are based on simulations for 6 fairness groups
% % and at least 1000 samples per group,
% % and we increase them slightly to accommodate for potential variations
% % of the content
% % and recording context across different languages in the database.
% We measure the difference in mean value and
% the relative difference per bin for regression.
% % For the test metric relative difference per bin
% % we require at least 67 predictions per bin in the combined test set,
% % or the bin is skipped.
% For classification, we measure the relative difference per class.

% --- Fairness Linguistic Sentiment
It has been shown that pre-trained transformer models
can use linguistic information
to improve their predictions~\citep{triantafyllopoulos2022}.
The \textbf{Fairness Linguistic Sentiment} tests
investigate the effect of linguistic content
for different types of languages.
If the textual content does have an influence on the model predictions,
it should have the same influence for each language on a fair model.
To this end,
we follow~\citet{triantafyllopoulos2022}
and employ \textsc{CheckList}~\citep{checklist2020},
a toolkit for generating automatic tests for \ac{NLP} models,
including sentiment models.
We expand on this by not only synthesising the English sentiment testing suite,
but also generating translated versions
of the test sentences for the languages German,
English,
Spanish,
French,
Italian,
Japanese,
Portuguese,
and Mandarin Chinese.
We use an \textsc{Opus-MT}~\citep{tiedemannThottingal2020}
model\footnote{https://huggingface.co/Helsinki-NLP/opus-mt-en-zh, accessed 2023/12/04}
for translation of Mandarin Chinese
and \textsc{Argos Translate}~\citep{argostranslate}
for all other languages,
followed by manual editing
to correct obvious translation errors. %BS: can you provide more info on the version of these models?
% HW:
% * version for argos is given in reference
% * the Opus-MT models are from Hugging Face,
%   and do not directly have versions attached to them, e.g.
%   https://huggingface.co/Helsinki-NLP/opus-mt-en-ar.
%   The original repo with the models is available at:
%   https://github.com/Helsinki-NLP/Opus-MT-train/tree/master/models
For the synthesis of each language,
a publicly available text-to-speech model % @felix: i guess you meant text-to-speech?
%AD corrected
using both the libraries \textsc{Coqui TTS}~\citep{coquitts}
% coqui TTS version is given in reference
and \textsc{ESPnet}~\citep{hayashi2020espnet},
version 0.10.6,
generated the audio samples 
corresponding to the text.
We investigate the tests for negative, neutral, and positive words in context.
Up to 2000 random samples are selected for each test and each language.
The prediction of the combined data is then compared
to the prediction for each individual language.
In this test, we only measure the influence of text sentiment
for different languages
rather than general language biases,
which are addressed in the Fairness Language tests.
Therefore,
we compare the shift in prediction
when filtering the samples for a specific sentiment.
We denote all samples with sentiment $s$ and language $l$ as $X_{l, s}$,
and all combined samples of language $l$ as $X_l$.
We compute the difference between the shift in prediction
for a certain sentiment and language
and the average of the shifts in prediction
for that sentiment
for all languages $l_i, 1 \leq i \leq L$

\begin{equation}
\label{eq:shiftdiff}
    \text{score}_{l, s} =
    \text{shift}(X_{l, s})
    - \frac{1}{L}\sum_{i=1}^{L} \text{shift}(X_{l_i,s}). 
\end{equation}

By subtracting the average shift in prediction for a certain sentiment,
we allow for both models
that are not affected by sentiment at all and models
that are affected by sentiment equally
across all languages to pass the tests.
For categorical emotion prediction,
we compute the shift in class proportion,
for negative, neutral, and positive sentiment.
For each class label $c$,
a fair model's behaviour in terms of class proportion shift
for one sentiment in one language should be similar
to the average behaviour observed across all languages.
This is tested by inserting the function $\text{shift}_c$,
which is defined as

\begin{align*}
    \text{shift}_{c}(X_{l,s}) =
    &\frac{1}{| X_{l,s} |} | \{ y \, | \, y = c \text{ and } y \in \text{prediction}(X_{l, s}) \}| \\
    &- \frac{1}{| X_l |} | \{ y \,| \, y = c \text{ and } y \in \text{prediction}(X_l) \}|,
\end{align*}

\noindent
in Eq.~\ref{eq:shiftdiff}.
$\text{shift}_c$ computes the proportion of samples
that were predicted as class $c$
among all samples with language $l$ and sentiment $s$
and subtracts the proportion of samples predicted as class $c$
among all samples of language $l$ (including all sentiments).
We apply the same function for dimensional emotion
values, by binning the model outputs and treating
the four bins as classes $c$.
We then compute the difference in bin proportion shift
analogously to the difference in class proportion shift.
Additionally, we consider the shift in terms of mean value
for negative, neutral, and positive sentiment.
Specifically,
we insert the following function
$\text{shift}_\text{mean}$ in Eq.~\ref{eq:shiftdiff}:

\begin{align*}
    \text{shift}_{\text{mean}}(X_{l, s}) = 
    &\text{mean}(\text{prediction}(X_{l, s})) \\
    & - \text{mean}(\text{prediction}(X_l)).
\end{align*}

We thus compute the difference between the shift in mean value
for one language
and the average shift in mean value across all languages.
% AD: Combined with class proportion shift function to shorten text
% Secondly,
% we apply Eq.~\ref{eq:shiftdiff} by inserting the shift
% in terms of bin proportions,
% $\text{shift}_b$,
% which is defined analogously to $\text{shift}_c$ for categories,

% \begin{align*}
%     \text{shift}_{b}(X_{l,s}) =
%     &\frac{1}{| X_{l,s} |} | \{ y \, | \, y = b \text{ and } y \in \text{prediction}_\text{bin}(X_{l, s}) \}| \\
%     &- \frac{1}{| X_l |} | \{ y \,| \, y = b \text{ and } y \in \text{prediction}_\text{bin}(X_l) \}|,
% \end{align*}

% where $b$ is the tested bin
% and $\text{prediction}_{\text{bin}}$ is a function
% that applies the model to a given set of samples
% and assigns one of four bin labels to each of the model outputs.
% We then measure the difference
% between the shift in bin proportion for one language
% and the average shift in bin proportion across all languages.
% The listed thresholds are based on simulations
% with random models for 8 fairness groups and at least 1000 samples per group.
% For the bin proportion shift difference metric
% we require at least 67 predictions per bin in the combined test set per sentiment,
% or we skip the bin for that sentiment.
%BS: perhaps all these thresholds and minima of samples could simply go into a table? Reading this is quite tiring ;)

% --- Fairness Pitch
The \textbf{Fairness Pitch} tests
address the different levels of average pitch a speaker can have.
Pitch is known to be correlated with emotion,
for instance, it has been observed
that a higher pitch leads to higher arousal~\citep{jaiswal2021robustness}.
An \ac{SER} model might use this correlation as a shortcut in its deductions,
leading to a disparate treatment of speakers in certain pitch ranges.
Consequently, we check the model behaviour for speakers
of different average pitch groups on data
with ground truth emotion labels
and exclude speakers with fewer than 25 samples.
For both categories and dimensions,
we use the \acs{msppodcast} test set 1.
We extract F0 frame-wise with \textsc{praat}~\citep{boersma2023praat}
and calculate a mean value for each segment,
ignoring frames with a pitch value of \SI{0}{Hz}.
We exclude segments from the analysis that show an F0 below \SI{50}{Hz}
or above \SI{350}{Hz}
to avoid pitch estimation outliers to influence the tests.
We then compute the average of all samples
belonging to a speaker,
and assign one of 3 pitch groups to that speaker.
The low pitch group is assigned to speakers
with an average pitch less than or equal to \SI{145}{Hz},
the medium pitch group to speakers with an average pitch of more than \SI{145}{Hz}
but less than or equal to \SI{190}{Hz},
and the high pitch group to speakers
with an average pitch higher than \SI{190}{Hz}.
We compute the performance of each pitch group
and compare it to the performance of the combined dataset.
% For regression, we measure the difference in
% \ac{CCC}, \ac{MAE}, \ac{MDE}, precision per bin, and recall per bin,
% and for classification the difference in
% \ac{UAR}, \ac{UAP}, \ac{RPC}, and \ac{PPC}.
%BS: this is becoming a) ultra-repetitive (table?!), b) a lot of "random" (links to repository to reproduce) and c) massively arbitrary (a lot of random numbers like "67" samples - why? where do these come from - more summarisation and reasoning could be good :) ) Perhaps even move some of this to an appendix? 

% --- Fairness Sex
For the \textbf{Fairness Sex} tests,
we select test sets
that have been labelled for the emotion task
as well as for sex
% and include a high number of speakers
% and a high number of samples per speaker.
% We calculate the performance for female
% (or male)
% speakers
and compute the difference to the performance
of the combined dataset.
% As with the Fairness Pitch test,
% we measure difference in \ac{CCC},
% \ac{MAE}, \ac{MDE}, precision per bin, and recall per bin for regression
% and \ac{UAR}, \ac{UAP}, \ac{RPC}, and \ac{PPC} for classification.

%==========================================================================
\subsection{Robustness Tests}

A robust machine learning model is resilient
when facing perturbations in the input data.
Robustness can be evaluated
by analysing how much the model predictions are affected
by changes such as noise.
The subcategory of adversarial robustness
specifically deals with perturbations
that are designed to be hard to detect
and change the model's prediction
(adversarial examples)~\citep{szegedy2013intriguing}.
We focus on applying perturbations
that are likely to occur
for non-malicious application scenarios
rather than adversarial attacks.

When ground truth labels are available,
one way to evaluate robustness
is to check the difference in correctness
with and without added noise~\citep{zhang2020}.
For regression, we check the difference in \ac{CCC}
and for classification the difference in \ac{UAR}.
% AD: removed metric
% and \ac{UAP}.
Another evaluation metric
is to consider
how often a perturbation changes the output,
which is presented as \emph{adversarial frequency}
in \citet{bastani2016measuring}.
We base our metric \emph{percentage of unchanged predictions} on this concept.
For classification,
the percentage of unchanged predictions
is simply the percentage of samples
where the class label prediction
does not change from the clean audio to the audio with perturbation.
It is not as straightforward to define
what counts as an unchanged prediction for continuous values.
For our tests, we set a threshold of $0.05$,
i.\,e., two predictions are considered to be unchanged
if their absolute difference is below $0.05$.
This metric can be used for labelled
as well as for unlabelled datasets.
% AD: removed metric
% For regression, we additionally consider the change in average value
% between the clean audio
% and the audio with perturbations.

%AD: adjusted table by removing metrics:
% Change UAP, Change Average Value
\begin{table*}[t]
	\centering
	\caption{Overview of the robustness tests, their test sets, metrics, and passing conditions.}
	\label{tab:robustness_thresholds}
\begin{tabular}{p{0.2\textwidth}p{0.075\textwidth}p{0.3\textwidth}p{0.2\textwidth}p{0.075\textwidth}}
\toprule
\textbf{Test} & \textbf{Task} & \textbf{Test Sets} & \textbf{Metric} & \textbf{Condition}                           \\
\midrule
Robustness Background Noise & categories & \acs{crema-d}, \acs{emovo}, \acs{iemocap}, \acs{meld},  & Change UAR &                   $>-0.05$ \\
                         &            & \acs{msppodcast} (test-1) & Perc. Unchanged Predictions &                     $>0.9$ \\
                         & dimensions & \acs{iemocap}, \acs{msppodcast} (test-1) & Change CCC &                   $>-0.05$ \\
                         &            &                                          & Perc. Unchanged Predictions &                     $>0.9$ \\
Robustness Low Quality Phone & categories & \acs{crema-d}, \acs{emovo}, \acs{iemocap}, \acs{meld}, & Change UAR &                   $>-0.05$ \\
                         &            &  \acs{msppodcast} (test-1) & Perc. Unchanged Predictions &                     $>0.5$ \\
                         & dimensions & \acs{iemocap}, \acs{msppodcast} (test-1) & Change CCC &                   $>-0.05$ \\
                         &            &                                          & Perc. Unchanged Predictions &                     $>0.5$ \\
Robustness Rec. Condition & categories, dimensions & \acs{nsc} & Perc. Unchanged Predictions &                     $>0.8$ \\
Robustness Sim. Rec. Condition & categories, dimensions & \acs{emovo}, \acs{nsc}, \acs{timit} & Perc. Unchanged Predictions &                     $>0.8$ \\
Robustness Small Changes & categories & \acs{crema-d}, \acs{emovo}, \acs{iemocap}, \acs{meld}, \acs{msppodcast} (test-1) & Perc. Unchanged Predictions &                    $>0.95$ \\
                         & dimensions & \acs{iemocap}, \acs{msppodcast} (test-1) & Perc. Unchanged Predictions &                    $>0.95$ \\
Robustness Spectral Tilt & categories & \acs{crema-d}, \acs{emovo}, \acs{iemocap}, \acs{meld}, & Change UAR &                   $>-0.02$ \\
                         &            & \acs{msppodcast} (test-1) & Perc. Unchanged Predictions &                     $>0.8$ \\
                         & dimensions & \acs{iemocap}, \acs{msppodcast} (test-1) & Change CCC &                   $>-0.05$ \\
                         &            &                                          & Perc. Unchanged Predictions &                     $>0.8$ \\
\bottomrule
\end{tabular}
\end{table*}

\cref{tab:robustness_thresholds} gives an overview of the discussed robustness tests,
their test sets, and suggested passing conditions if a binary test result is desired.
As with the correctness tests, the thresholds for the passing conditions are application
dependent, and the thresholds in the table are only an example
to show how the test results can be summarised into a percentage of passed tests.
%AD: adjusted number for reduced metrics
% The number of robustness tests are 82
% each for arousal, dominance, and valence, 
% and 193 for categorical emotions.
The number of robustness tests are 64
each for arousal, dominance, and valence, 
and 148 for categorical emotions.
In the following paragraphs,
we provide additional information to the single tests,
besides what is shown in \cref{tab:robustness_thresholds}.

% --- Robustness Background Noise
\ac{SER} models have been shown to be affected by background noises
although the human perception of the emotion
remains the same \cite{jaiswal2021robustness}.
The \textbf{Robustness Background Noise} tests
investigate the robustness for various types of background noises.
We simulate babble noise by mixing 4-7 speech samples
from \ac{musan}~\citep{musan2015}
and adding them
with a \ac{SNR} of \SI{20}{dB}.
\ac{musan} also contains technical and ambient sounds,
as well as a music,
from which one sample is added with an \ac{SNR} of \SI{20}{dB}
for simulating environmental noise
and music, respectively.
We also check the effect of human coughing and sneezing sounds
with samples collected by \citet{Amiriparian2017}
by adding a single cough or sneeze
at a random position with an \ac{SNR} of \SI{10}{dB}.
For artificial noise,
we add white noise with an \ac{SNR} of \SI{20}{dB}.
% For each type of background noise,
% measure the percentage of unchanged predictions, the change in average value, \ac{CCC}, \ac{UAR}, and \ac{UAP}.

% --- Robustness Low Quality Phone
The \textbf{Robustness Low Quality Phone} tests
specifically target applications
with audio from a low quality telephone connection.
These types of recordings usually display a stronger compression,
coding artifacts,
and may show low pass behaviour.
We mimic this by applying a dynamic range compressor,
a lossy (narrow band) \ac{AMR} codec,
and high pass filtered pink noise.
% We measure the percentage of unchanged predictions, the change in average value, \ac{CCC}, \ac{UAR}, and \ac{UAP}.
% For the percentage of unchanged predictions
% and the change in average value, 
% we select less strict thresholds with $0.5$
% and $0.05$, respectively,
% as this perturbation has a higher impact on the perception of the audio.

% --- Robustness Recording Condition
Databases such as the \ac{nsc}~\citep{koh19building}
contain multiple samples
of the same audio recorded simultaneously
with different recording devices.
We use this data
in the \textbf{Robustness Recording Condition} tests
and compare the predictions of the baseline recording device
to the predictions of audio from alternative devices.
In the case of the \ac{nsc} dataset,
we randomly select 5000 samples
from the headset recordings
and compare them to their respective recordings
using the boundary microphone,
as well as to their respective mobile phone recordings,
and compute the percentage of unchanged predictions.

% --- Robustness Simulated Recording Condition
Another option to evaluate robustness
for different recording conditions
is to simulate them.
The \textbf{Robustness Simulated Recording Condition} tests
simulate audio recordings at different locations
using impulse responses
from the \ac{mardy}~\citep{wen2006evaluation} dataset,
and different rooms using impulse responses
from the \ac{air}~\citep{jeub09a} dataset.
We use the impulse response
in the centre position at 1 meter distance
as the baseline
to test robustness to other positions.
For the room test, 
we use the impulse response of a recording booth as reference
and compare to impulse responses of other rooms
recorded at similar distances as the reference.
We apply the impulse responses to \ac{emovo},
to 5000 randomly selected headset recordings
from the \ac{nsc} dataset,
and to 5000 randomly selected samples
from the \ac{timit}~\citep{timit},
which contains broadband recordings of 630 speakers
reading ten sentences in American English.
We select those three datasets
as they provide dry speech recordings
with a high \ac{SNR}.
Then, we measure the percentage of unchanged predictions.

% --- Robustness Small Changes
The \textbf{Robustness Small Changes} tests
apply very small transformations to the audio
that were designed to be perceived as subtle to a person.
Each of the small augmentations
described in the following
is compared to the baseline.
The Additive Tone test adds a sinusoid
with a randomly selected frequency between \SI{5000}{Hz} and \SI{7000}{Hz}
with an \ac{SNR} of \SI{40}{dB}, \SI{45}{db}, or \SI{50}{dB}.
The Append Zeros and Prepend Zeros tests
add $100$, $500$, or $1000$ samples containing zeros
to the end of the signal,
or respectively the beginning of the signal.
The Clip test clips $0.1\%$, $0.2\%$, or $0.3\%$  of the input sample.
The Crop Beginning and Crop End tests
remove $100$, $500$, or $1000$ samples
from the beginning or end of the signal respectively.
The Gain test changes the gain of the signal
by a value randomly selected
from \SI{-2}{dB}, \SI{-1}{dB}, \SI{1}{dB}, and \SI{2}{dB}.
The Highpass Filter and Lowpass Filter tests
apply a Butterworth filter of order $1$
with a cutoff frequency
of \SI{50}{Hz}, \SI{100}{Hz}, or \SI{150}{Hz} for a lowpass, or \SI{7500}{Hz}, 
\SI{7000}{Hz}, or \SI{6500}{Hz} for a highpass.
The White Noise test adds Gaussian distributed white noise
with a root mean square based \ac{SNR}
randomly selected from \SI{35}{dB}, \SI{40}{dB}, and \SI{45}{dB}.
% For each of the described small augmentations,
% we measure the percentage of unchanged predictions.

% --- Robustness Spectral Tilt
The \textbf{Robustness Spectral Tilt} tests
simulate the boosting of low or high frequencies in the spectrum.
We simulate such spectral tilts
by attenuating or emphasising the signal linearly,
while ensuring that the overall signal level stays the same
if possible without clipping.
% We measure the percentage of unchanged predictions, the change in average value, \ac{CCC}, \ac{UAR}, and \ac{UAP}.

%%%%%%%%%%%%%%%%%%%%%%%%%%%%%%%%%%%%%%%%%%%%%%%%%%%%%%%%%%%%%%%%%%%%%%%%%%%
\section{Models}

\addtolength{\tabcolsep}{-0.3em}  % Reduce space bewteen columns
\begin{table}[t]
    \centering
    \caption{
    Tested models, their architecture, number of layers, number of parameters, datasets (hours of speech), and languages during pre-training.
    The number of languages for \emph{axlstm}
    where estimated with a whisper-large-v3 model.
    }
    \begin{tabular}{llrrll}
        \toprule
        \textbf{Model} & \textbf{Archit.} & \textbf{Layer} & \textbf{Param.} & \textbf{Datasets} & \textbf{Lang.} \\
        \midrule
        wavlm         & Transf. & 24 & 319\,M & \makecell[lt]{Libri-Light (60\,k)\\GigaSpeech (10\,k)\\VoxPopuli (24\,k)} & eng \\
        data2vec      & Transf. & 24 & 314\,M & LibriSpeech (960) & eng \\
        CNN14         & CNN & 14         &  80\,M & -           &  - \\
        hubert-b      & Transf. & 12 &  95\,M & LibriSpeech (960) & eng \\
        w2v2-b        & Transf. & 12 &  95\,M & LibriSpeech (960) &  eng \\
        hubert-L      & Transf. & 24 & 316\,M & Libri-Light (60\,k) & eng \\
        w2v2-L        & Transf. & 24 & 316\,M & LibriSpeech (960) & eng \\
        w2v2-L-robust & Transf. & 24 & 316\,M & \makecell[lt]{Libri-Light (60\,k)\\Fisher (2\,k)\\CommonVoice (700)\\Switchboard (300)} & eng \\
        w2v2-L-vox    & Transf. & 24 & 316\,M & VoxPopuli (100\,k) & 23 \\
        w2v2-L-xls-r  & Transf. & 24 & 316\,M & \makecell[lt]{VoxPopuli (372\,k)\\MLS (50\,k)\\CommonVoice (7\,k)\\VoxLingua107 (6.6\,k)\\BABEL (1\,k)} & 128 \\
        axlstm        & xLSTM & 12       &  44\,M &   AudioSet (3\,k) &   $>$80 \\
        \bottomrule
    \end{tabular}
    \label{tab:models}
\end{table}
\addtolength{\tabcolsep}{0.3em}  % Restore space between columns

\cref{tab:models} lists the eleven tested models,
summarising their size in parameters,
hours of speech and languages included during pre-training.
The \emph{wavlm} model~\citep{chen2022wavlm} has 24 transformer layers, and was pre-trained on English data
(Libri-Light~\citep{librilight2020},
10\,k hour subset of GigaSpeech~\citep{chen2021gigaspeech},
24\,k hours English subset of VoxPopuli~\citep{voxpopuli2021}).
The \emph{data2vec} model~\citep{baevski2022data2vec}
consists of 24 transformer layers,
and was pre-trained on English data (LibriSpeech corpus~\citep{librispeech2015}).
We further test all models trained by \citet{wagner2023dawn},
namely a 14-layer Convolutional Neural Network (\emph{CNN14}),
which was not pre-trained;
four models based on the HuBERT~\citep{hsu2021hubert}
and wav2vec 2.0~\citep{baevski2020wav2vec} architectures,
each pre-trained on English audiobooks,
using 12 transformer layers (\emph{hubert-b} and \emph{w2v2-b}),
or 24 transformer layers (\emph{hubert-L} and \emph{w2v2-L}),
where \emph{hubert-L} has been pre-trained
on the Libri-Light corpus~\citep{librilight2020}
and the other three have been pre-trained
on the LibriSpeech corpus~\citep{librispeech2015};
\emph{w2v2-L-robust},
a model identical to \emph{w2v2-L},
but instead pre-trained on English recordings
of audiobooks (Libri-Light~\citep{librilight2020}),
Wikipedia sentences (Common Voice~\citep{commonvoice2020}),
and telephone speech
(Switchboard~\citep{switchboard1993} and Fisher~\citep{fisher2004});
\emph{w2v2-L-vox},
a model identical to \emph{w2v2-L}
but instead pre-trained on parliamentary speech in multiple languages 
(VoxPopuli~\citep{voxpopuli2021});
and \emph{w2v2-L-xls-r},
a model identical to \emph{w2v2-L}
but instead pre-trained on more than 400\,k hours across all domains
and multiple languages
(VoxPopuli~\citep{voxpopuli2021},
MLS~\citep{mllibrispeech2020},
Common Voice~\citep{commonvoice2020},
VoxLingua107~\citep{voxlingua107},
and BABEL~\citep{babel2014}).
We test
the \emph{axlstm} model~\citep{yadav2024axlstm}
as an alternative to transformer based models.
We use the \emph{axlstm base} model,
which has 44\,M parameters
and was pre-trained on 5\,k hours of different audio data from AudioSet~\citep{gemmeke2017audioset},
including 3\,k of speech.

We fine-tuned all models
on the MSP-Podcast corpus~(v1.7)~\citep{lotfian2019msppodcast}
with the multitask of arousal, dominance, and valence.
For each,
we also train a categorical counterpart with the four classes
anger, happiness, neutral, and sadness.
We start with the same pre-trained models
and fine-tune them on the same \acs{msppodcast} data,
but use the categorical labels as targets instead. 
All transformer based models are fine-tuned
using a single random seed,
the ADAM optimiser with CCC loss for dimensions
and with weighted cross-entropy loss for categories.
The learning rate is fixed to $1e-4$ for all models
except for \emph{data2vec} and \emph{w2v2-L} (categorical),
which failed to learn with that learning rate,
and are thus trained with a learning rate of $5e-5$.
We run for 5 epochs with an effective batch size of 32
and keep the checkpoint with the best performance
on the development set.
During fine-tuning, we freeze the CNN layers
but fine-tune the transformer ones.
An exception is \emph{wavlm},
which we did not fine-tune ourselves
but used the model fine-tuned on MSP-Podcast~(v1.11)~\citep{lotfian2019msppodcast}
with attentive statistics pooling
from the MSP-Podcast challenge 2024~\citep{goncalves2024mspchallenge},
whereas for categorical emotions,
we ignore the classes surprise, fear, disgust, and contempt.
The \emph{CNN14}~\citep{kong2020panns} model is trained for 60 epochs
with a learning rate of $1e-3$, a batch size of 16,
and for 5 random seeds.
For the \emph{axlstm} model,
we follow the fine-tuning approach in~\citep{yadav2024axlstm}
and freeze the whole pre-trained model,
and use its fixed sized feature vectors
to train a single hidden layer
MLP classifier with 1024 neurons,
a batch size of 1024,
and a learning rate of $1e-4$, 
for 500 epochs.

%%%%%%%%%%%%%%%%%%%%%%%%%%%%%%%%%%%%%%%%%%%%%%%%%%%%%%%%%%%%%%%%%%%%%%%%%%%
\section{Results}
\label{sec:results}

\begin{table}[t]
    \centering
    \caption{Percentage of passed tests
    for each of the emotion prediction tasks
    (\textbf{A}rousal,
    \textbf{D}ominance,
    \textbf{V}alence,
    Emotionial \textbf{C}ategories. % @felix: i find C for categories more intuitive than E
    % HW: changed accordingly
    % AD: changed to address "Table 5 should include that the models are listed based on their average performance."
    The last column presents the average (\textbf{$\varnothing$})
    over the four values
    and the models are ranked based on this average.
    The model with the most passed tests
    is marked in bold in each column.
    % and as average $\varnothing$ over those four.
    %The models are ranked by the average
    %and the winning value is marked in bold.
    % The number of tests are
    % 727 (66/557/82) for \textbf{A},
    % 711 (72/557/82) for \textbf{D},
    % 715 (76/557/82) for \textbf{V},
    % 909 (196/520/193) for \textbf{C},
    % with the number of correctness/fairness/robustness tests shown in parentheses.
    % HW: the number of tests is now stated in the method section
    }
    \label{tab:results-overview}
    \begin{tabular}{lrrrrr}
        \hline \\ [-2ex]
        \textbf{Model} & \textbf{A} & \textbf{D} & \textbf{V} & \textbf{C} & \textbf{$\varnothing$} \\
        \hline \\ [-2ex]
        \multicolumn{6}{l}{\textbf{All tests}} \\

        % Results of original submission
        %w2v2-L-robust & .871 & \textbf{.854} & .805 & .771 & \textbf{.825} \\
        %hubert-L      & .859 & .847 & .794 & \textbf{.785} & .821 \\
        %w2v2-L-vox    & \textbf{.875} & .844 & .811 & .750 & .820 \\
        %hubert-b      & .866 & .853 & \textbf{.812} & .724 & .814 \\
        %w2v2-L-xls-r  & .854 & .842 & .776 & .764 & .809 \\
        %w2v2-L        & .870 & .817 & .794 & .745 & .806 \\
        %w2v2-b        & .837 & .839 & .778 & .743 & .799 \\
        %CNN14         & .796 & .815 & .739 & .727 & .769 \\

        % Results for re-submission, using fewer test cases
        wavlm         & \textbf{.883} & .851 & .834 & .785 & \textbf{.838} \\
        w2v2-L-vox    & \textbf{.883} & .851 & \textbf{.837} & .763 & .834 \\
        w2v2-L-robust & .874 & .844 & .809 & .789 & .829 \\
        hubert-L      & .864 & .847 & .788 & \textbf{.795} & .824 \\
        hubert-b      & .871 & \textbf{.854} & .818 & .730 & .818 \\
        w2v2-L-xls-r  & .860 & .836 & .804 & .767 & .817 \\
        w2v2-L        & .879 & .817 & .810 & .745 & .813 \\
        w2v2-b        & .849 & .849 & .792 & .753 & .811 \\
        data2vec      & .845 & .821 & .769 & .743 & .794 \\
        CNN14         & .805 & .832 & .792 & .748 & .794 \\
        axlstm        & .837 & .825 & .794 & .714 & .792 \\

        \hline \\ [-2ex]
        \multicolumn{6}{l}{\textbf{Correctness tests}} \\

        % Results of original submission
        %hubert-L      & \textbf{.890} & .730 & .583 & \textbf{.571} & \textbf{.694} \\
        %w2v2-L-robust & .841 & .769 & \textbf{.596} & .554 & .690 \\
        %w2v2-L        & .880 & \textbf{.787} & .436 & .488 & .648 \\
        %w2v2-L-xls-r  & .835 & .782 & .411 & .382 & .602 \\
        %hubert-b      & .741 & .701 & .492 & .452 & .596 \\
        %w2v2-b        & .689 & .719 & .377 & .493 & .570 \\
        %w2v2-L-vox    & .758 & .752 & .361 & .386 & .564 \\
        %CNN14         & .625 & .515 & .281 & .390 & .453 \\

        % Results for re-submission, using fewer test cases
        hubert-L      & \textbf{.857} & .747 & .550 & \textbf{.639} & \textbf{.698} \\    
        w2v2-L-robust & .841 & .769 & .563 & .578 & .688 \\
        data2vec      & .797 & .764 & .560 & .542 & .666 \\
        wavlm         & .652 & .708 & \textbf{.566} & .619 & .636 \\
        w2v2-L        & .846 & \textbf{.778} & .386 & .513 & .631 \\
        hubert-b      & .708 & .760 & .517 & .470 & .614 \\
        w2v2-b        & .714 & .769 & .360 & .589 & .608 \\
        w2v2-L-xls-r  & .802 & .774 & .386 & .395 & .589 \\
        w2v2-L-vox    & .724 & .769 & .403 & .414 & .578 \\
        axlstm        & .730 & .725 & .243 & .477 & .544 \\
        CNN14         & .675 & .548 & .347 & .424 & .498 \\ 

        \hline \\ [-2ex]
        \multicolumn{6}{l}{\textbf{Fairness tests}} \\

        % Results of original submission
        %w2v2-L-vox    & .947 & .945 & .982 & .948 & \textbf{.955} \\
        %hubert-b      & .958 & \textbf{.970} & .972 & .911 & .953 \\
        %w2v2-L-xls-r  & .944 & .939 & .943 & .976 & .950 \\
        %w2v2-L        & \textbf{.964} & .900 & \textbf{.994} & .917 & .944 \\
        %hubert-L      & .917 & .953 & .944 & .941 & .939 \\
        %w2v2-L-robust & .915 & .923 & .930 & .952 & .930 \\
        %w2v2-b        & .921 & .921 & .927 & .948 & .929 \\        
        %CNN14         & .942 & .933 & .844 & \textbf{.988} & .927 \\

        % Results for re-submission, using fewer test cases
        axlstm        & \textbf{.979} & .964 & \textbf{1.000} & .987 & \textbf{.982} \\
        w2v2-L-vox    & .957 & .943 & .978 & .954 & .958 \\
        hubert-b      & .968 & \textbf{.970} & .972 & .911 & .955 \\
        w2v2-L-xls-r  & .951 & .934 & .954 & .976 & .954 \\
        CNN14         & .952 & .930 & .919 & \textbf{.988} & .947 \\ 
        w2v2-L        & .973 & .895 & .994 & .916 & .944 \\
        hubert-L      & .909 & .952 & .944 & .947 & .938 \\
        w2v2-b        & .936 & .936 & .927 & .948 & .937 \\
        w2v2-L-robust & .909 & .923 & .930 & .952 & .928 \\
        wavlm         & .940 & .915 & .946 & .902 & .926 \\
        data2vec      & .942 & .903 & .880 & .932 & .914 \\

        \hline \\ [-2ex]
        \multicolumn{6}{l}{\textbf{Robustness tests}} \\

        % Results of original submission
        %w2v2-L-robust & \textbf{.714} & .\textbf{788} & .567 & .750 & \textbf{.705} \\
        %hubert-L      & .613 & .668 & \textbf{.598} & .796 & .669 \\
        %w2v2-L-vox    & .705 & .580 & .430 & \textbf{.822} & .634 \\
        %hubert-b      & .434 & .582 & .395 & .593 & .501 \\
        %w2v2-L-xls-r  & .400 & .534 & .364 & .704 & .500 \\
        %w2v2-L        & .442 & .483 & .462 & .562 & .487 \\
        %w2v2-b        & .516 & .469 & .310 & .480 & .444 \\
        %CNN14         & .259 & .282 & .209 & .467 & .304 \\

        % Results for re-submission, using fewer test cases
        w2v2-L-robust & \textbf{.728} & \textbf{.783} & .532 & .731 & \textbf{.694} \\
        wavlm         & .701 & .646 & .628 & .703 & .670 \\
        w2v2-L-vox    & .706 & .671 & .474 & \textbf{.803} & .664 \\
        hubert-L      & .643 & .652 & .554 & .769 & .654 \\
        data2vec      & .503 & .581 & \textbf{.649} & .579 & .578 \\
        w2v2-L-xls-r  & .344 & .497 & .413 & .658 & .478 \\
        w2v2-L        & .414 & .449 & .439 & .521 & .456 \\
        w2v2-b        & .511 & .504 & .299 & .480 & .448 \\
        hubert-b      & .370 & .474 & .322 & .540 & .426 \\
        CNN14         & .250 & .293 & .207 & .476 & .306 \\
        axlstm        & .199 & .199 & .280 & .353 & .258 \\

        \hline
    \end{tabular}
\end{table}

%BS: please make sure to reference every table and figure also in the text
None of the models
passes all tests
as indicated by the percentage of passed tests
in \cref{tab:results-overview}
for the four different tasks
arousal, dominance, valence, and emotional categories.
The percentage of passed test results
are computed as the average percentage
over the involved tests,
to better reflect that single tests differ
in the number of involved metrics and test sets.

The test results allow a comparison
of the different models.
The models can also be ranked
based on the percentage of passed tests,
but this has to be done carefully
as different tests might have different importance
for particular applications.
Further, 
the single tests,
or tests grouped by correctness, fairness, robustness
can provide insights into model behaviour.

\emph{wavlm} passes the most tests
for all tasks.
All models pre-trained on speech
are ranked before the \emph{CNN14} baseline
and the \emph{axlstm} model,
pre-trained on AudioSet.

Detailed results with additional plots
are available under \url{https://audeering.github.io/ser-tests/}.
% Update page on github!!!
In the following sub-sections,
we focus on a few interesting results.

%==========================================================================
\subsection{Correctness}

\begin{figure}[t]
    \centering
    \includegraphics[width=\columnwidth]{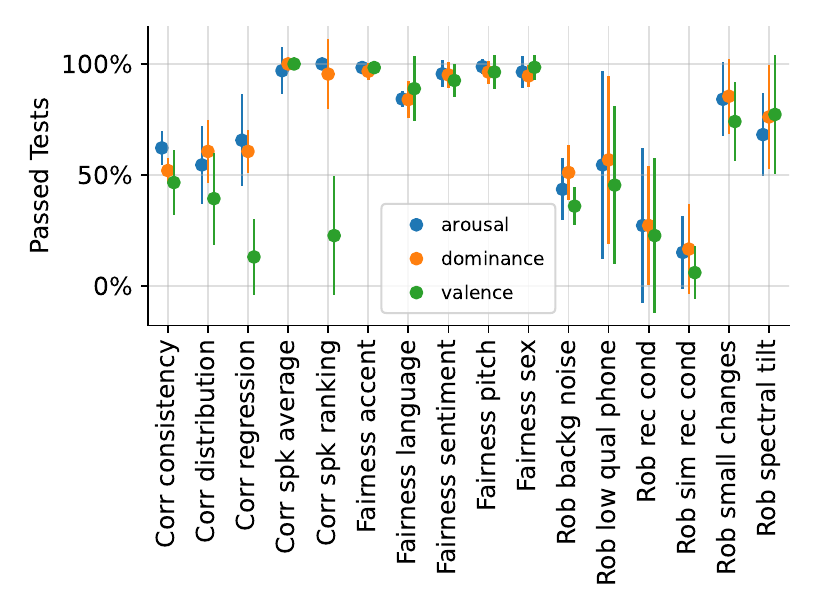}
    \caption{
        Percentage of passed tests averaged over all models
        presented with standard deviation
        for all tests involving a dimensional emotion task.
        Corr stands for Correctness,
        Rob for Robustness,
        spk for speaker,
        backg for background,
        qual for quality,
        rec for recording,
        and cond for condition.
    }
    \label{fig:dimensional}
\end{figure}

The test results for correctness
indicate
that the valence task is much harder for the models
(see \cref{fig:dimensional}).
For arousal and dominance, a model passes on average
$66$\% and $61$\% of the Correctness Regression tests,
for valence the average is only $13$\%.

% --- Correctness Speaker Rating/Average
Figure~\ref{fig:dimensional} shows a significant difference between
Correctness Speaker Average and Correctness Speaker Rating tests
for the valence task only.
Upon closer inspection of the test data,
this seems to be due to the fact
that the ground truth speaker averages lie more closely together
for valence than for the other tasks.

\begin{figure*}[t]
    \centering
    \includegraphics[width=2\columnwidth]{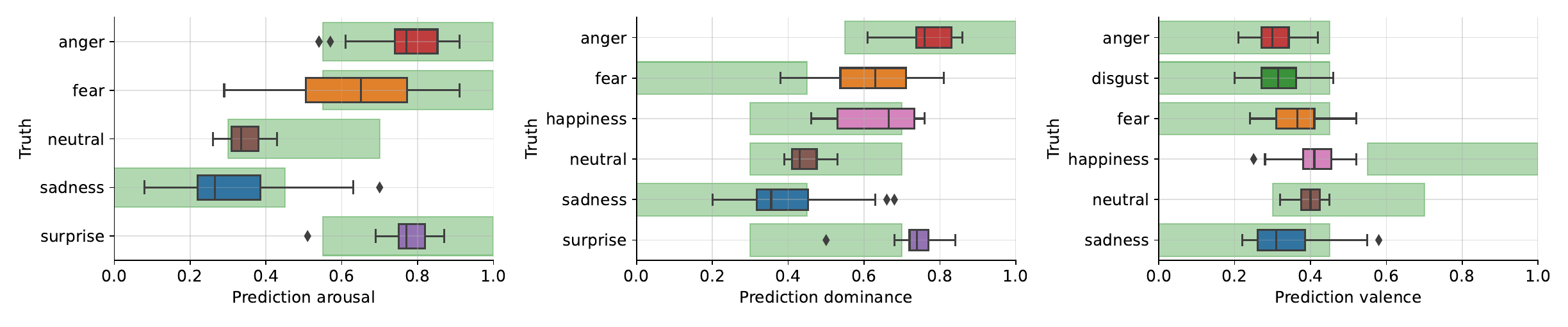}
    \caption{
        Predictions of \emph{hubert-L}
        for arousal (left),
        dominance (centre),
        valence (right)
        on the \ac{ravdess} test set,
        split by the categorical emotions
        the samples are annotated for in \ac{ravdess}.
        The green area marks the region
        in which a dimensional prediction
        would be rated as consistent
        with the annotated emotional category
        by the Correctness Consistency tests.
    }
    \label{fig:correctness-consistency}
\end{figure*}

% --- Correctness Consistency
The Correctness Consistency tests
estimate how well the models' arousal, dominance, and valence predictions
fit for a sample with an assigned emotional category as ground truth.
\emph{w2v2-L},
\emph{w2v2-L-vox},
\emph{w2v2-L-xls-r},
and \emph{wavlm}
are the most consistent.
\emph{hubert-L} and \emph{w2v2-L-robust}
pass a similar number of tests for arousal and dominance,
but are less consistent for valence,
where they tend to predict the same valence value
independent of the underlying categorical emotion.
Results for \emph{hubert-L} are shown in \cref{fig:correctness-consistency}.

%==========================================================================
\subsection{Fairness}
% AD: changed from 90 to 88, (data2vec has 88% for valence?)
All models pass at least $88$\%
of the fairness tests
for any task.
The models passing the most tests are
\emph{w2v2-L-vox},
\emph{hubert-b},
and \emph{w2v2-L-xls-r},
%AD: changed back from 88 to 96%
on average passing around $96$\% of the tests.
There is no single fairness test
that is completely failed by a model.
\emph{axlstm} passes the most fairness tests,
whereas \emph{data2vec} fails the most,
especially for the Fairness Language tests.

% --- Fairness Linguistic Sentiment
\emph{w2v2-L-vox} and \emph{axlstm} are the models least influenced
by linguistic sentiment,
both passing 99.4\% of the Fairness Linguistic Sentiment tests.
% It only shows a slight shift towards predicting the emotion \emph{anger}
% for Japanese and a neutral sentiment.
\emph{hubert-b},
\emph{w2v2-b},
\emph{hubert-L},
\emph{w2v2-L-robust},
\emph{wavlm},
and \emph{data2vec}
clearly show a strong influence by sentiment
when predicting valence and categorical emotion for English.
A positive sentiment leads to higher valence values,
a higher number of \emph{happiness}
and a reduced number of \emph{sadness} predictions,
and vice versa.
Interestingly,
the same trend is observable
with \emph{w2v2-L-robust},
\emph{hubert-L},
and \emph{data2vec}
for arousal and dominance as well.

%==========================================================================
\subsection{Robustness}

The robustness tests roughly group the models in three classes.
\emph{w2v2-L-robust},
\emph{wavlm},
\emph{w2v2-L-vox},
and \emph{hubert-L}
are more robust than all other models,
\emph{CNN14} and \emph{axlstm} are less robust
than all other models,
and the remaining models pass a similar number of robustness tests.
\emph{data2vec} is close to the first group
and shows highest robustness for valence,
but does not achieve a similar performance for arousal, dominance,
and emotional categories.
When comparing the models
\emph{data2vec},
\emph{w2v2-L},
\emph{hubert-b},
\emph{w2v2-b},
which are all pre-trained on the same dataset,
we see that larger models are more robust than smaller models.

\begin{figure*}[t]
    \centering
    \includegraphics[width=2\columnwidth]{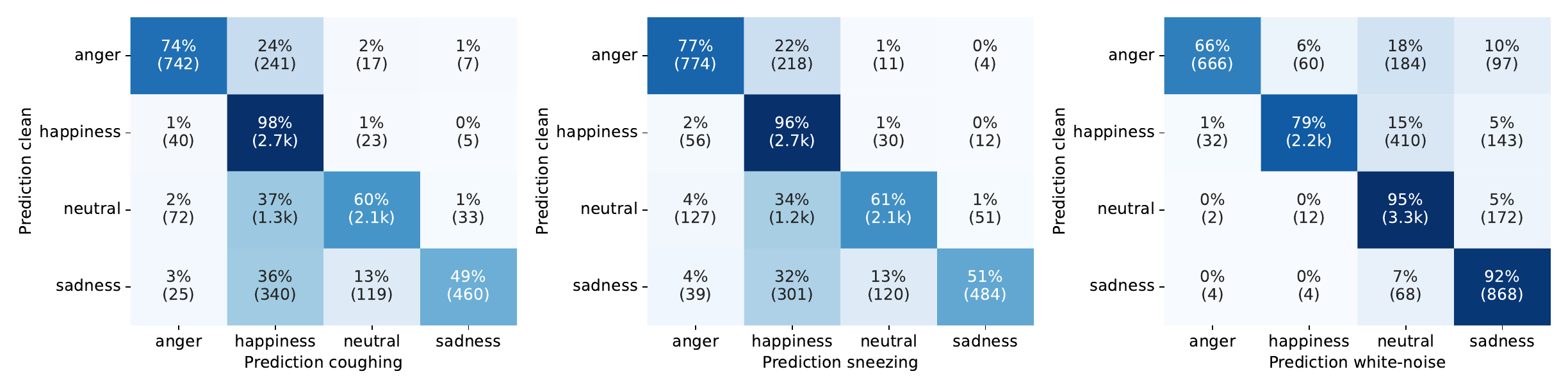}
    \caption{
        Confusion matrices for the prediction of emotional categories
        by \emph{hubert-L}
        on the clean \acs{msppodcast} test set 1
        comparing to the \acs{msppodcast} test set 1
        when adding coughing with an \ac{SNR} of \SI{10}{dB} (left),
        sneezing with an \ac{SNR} of \SI{10}{dB} (centre),
        or white noise with an \ac{SNR} of \SI{20}{dB} (right).
    }
    \label{fig:robustness-background-noise}
\end{figure*}

% --- Robustness Background Noise
Most of the models fail more than $50$\% of all Robustness Background Noise tests,
and are mainly affected by coughing,
sneezing,
and white noise.
% \emph{CNN14} is the most affected,
% as it shows failing tests
% for all types of background noise.
% For the emotional dimensions, 
% most of them show a tendency towards higher values
% when coughing or sneezing is present,
% lower values when white noise is present,
% and a drop in \ac{CCC} in all those cases.
% \emph{w2v2-L-robust} fails the fewest tests for
% arousal ($67$\%)
% and dominance ($78$\%).
% \emph{hubert-L} fails the fewest tests for valence ($47$\%)
% and \emph{w2v2-L-xls-r} for emotional categories ($72$\%).
When adding sneezing or coughing sounds
to the input signal
when predicting emotional categories,
the predictions of the models shifts
towards the \emph{happiness} class,
see \cref{fig:robustness-background-noise} for results for \emph{hubert-L},
indicating that coughing and sneezing might be confused
with laughter by the models.

%%%%%%%%%%%%%%%%%%%%%%%%%%%%%%%%%%%%%%%%%%%%%%%%%%%%%%%%%%%%%%%%%%%%%%%%%%%
\section{Discussion}

\begin{table}[t]
    \centering
    \caption{
        Average \ac{CCC} for arousal (\textbf{A}),
        dominance (\textbf{D}),
        and valence (\textbf{V}),
        and average \ac{UAR} for emotional categories (\textbf{C}).
        The last column presents the average (\textbf{$\varnothing$})
        over the four values
        and the models are ranked based on this average.
    }
    \label{tab:results-ccc}
    \begin{tabular}{llllll}
        \hline \\ [-2ex]
        \textbf{Model}
        & \textbf{A}
        & \textbf{D}
        & \textbf{V}
        & \textbf{C}
        & \textbf{$\varnothing$} \\
        \hline \\ [-2ex]

        wavlm         & .61 & \textbf{.54} & \textbf{.51} & \textbf{.64} & \textbf{.58} \\
        w2v2-L-robust & \textbf{.64} & \textbf{.54} & .50 & .58 & .57 \\
        hubert-L      & .63 & .53 & .49 & .55 & .55 \\
        data2vec      & .62 & .53 & .47 & .52 & .53 \\
        w2v2-L-vox    & .63 & \textbf{.54} & .36 & .50 & .51 \\
        hubert-b      & .60 & .52 & .40 & .51 & .51 \\
        w2v2-b        & .60 & .53 & .37 & .52 & .51 \\
        w2v2-L        & .63 & .53 & .32 & .53 & .50 \\
        w2v2-L-xls-r  & .63 & .53 & .30 & .50 & .49 \\
        axlstm        & .50 & .42 & .14 & .46 & .38 \\
        CNN14         & .50 & .39 & .18 & .44 & .38 \\
        
        \hline
    \end{tabular}
\end{table}

Machine learning models
for the speech emotion recognition tasks
are usually benchmarked based on their performance
in terms of \ac{CCC} for dimensional tasks
or \ac{UAR} for categorical tasks.
When calculating the average over \ac{CCC} and \ac{UAR}
of the models
over all four tasks,
\emph{wavlm} (.58)
and \emph{w2v2-L-robust} (.57)
are the best models
(see \cref{tab:results-ccc}).
%
% CCC values for arousal, dominance, valence; UAR for emotion
% | model         | arousal | dominance | valence | emotion | average |
% | ------------- | ------- | --------- | ------- | ------- | ------- |
% | w2v2-L-vox    | .63     | .54       | .36     | .50     | .51     |
% | w2v2-L-robust | .64     | .54       | .50     | .58     | .57     |
% | hubert-L      | .63     | .53       | .49     | .55     | .55     |
%
Interestingly,
the ranking is slightly different for the average number of passed correctness tests
in \cref{tab:results-overview}.
There \emph{wavlm} is ranked only fourth,
as it shows a low performance for the correctness tests on arousal.
This is mainly due to a shift towards negative in its predictions,
which only slightly affects the CCC value,
but is reflected in other test results like Correctness Distribution.
Looking at the overall number of passed tests,
\emph{wavlm} is again placed at the top of the rankings.

\begin{table}[t]
    \centering
    \caption{
        Change of the predicted average valence value
        for neutrally spoken words in English
        with \textbf{negative},
        \textbf{neutral},
        and \textbf{positive}
        sentiment.
    }
    \label{tab:results-sentiment}
    \begin{tabular}{llll}
        \hline \\ [-2ex]
        \textbf{Model} & \textbf{negative} & \textbf{neutral} & \textbf{positive} \\
        \hline \\ [-2ex]

        data2vec      & $-.15$   & $-.02$   & $+.16$   \\
        hubert-L      & $-.12$   & $+.02$   & $+.12$   \\
        wavlm         & $-.12$   & $\pm.00$ & $+.13$   \\
        w2v2-L-robust & $-.12$   & $\pm.00$ & $+.13$   \\
        hubert-b      & $-.09$   & $-.03$   & $+.11$   \\
        w2v2-b        & $-.03$   & $-.02$   & $+.05$   \\
        w2v2-L-vox    & $-.02$   & $\pm.00$ & $+.02$   \\
        w2v2-L        & $\pm.00$ & $+.01$   & $\pm.00$ \\
        w2v2-L-xls-r  & $\pm.00$ & $+.01$   & $\pm.00$ \\
        CNN14         & $-.01$   & $+.02$   & $\pm.00$ \\
        axlstm        & $-.01$   & $+.01$   & $+.01$   \\
        
        \hline
    \end{tabular}
\end{table}

\citet{triantafyllopoulos2022} showed
that some of the success in predicting valence
can be attributed to the linguistic knowledge
encoded in the self-attention layers
of transformer based models.
The results for the correctness tests for valence
(\cref{tab:results-overview})
and the \ac{CCC} values for valence
(\cref{tab:results-ccc})
indicate that
\emph{wavlm},
\emph{w2v2-L-robust},
\emph{hubert-L},
and \emph{data2vec}
are showing the best performance
and hence might
rely more on linguistic content.
This hypothesis can be checked
by evaluating how valence is influenced
by the sentiment of the spoken text.
This is measured as part of the
Fairness Linguistic Sentiment
test,
which synthesises neutrally spoken words/sentences
with negative, neutral, or positive sentiment.
\cref{tab:results-sentiment} lists
the shift in average predicted valence
for the different sentiment conditions
for English.
\emph{wavlm}, \emph{hubert-L}, \emph{data2vec}, and \emph{w2v2-L-robust},
are the models
that show the largest shift towards lower valence
for negative sentiment
and towards higher valence
for positive sentiment.
Whereas the models 
\emph{w2v2-L},
\emph{w2v2-L-xls-r},
and \emph{w2v2-L-vox}
show no, or only a small shift.
The results indicate
that the best performing models
for valence do indeed
take sentiment into account.
What seems to reduce the ability to learn linguistic information
is the inclusion of different languages
in the training data
as is the case for \emph{w2v2-L-xls-r}
which was trained on $\sim 7$ times the amount of data
than \emph{w2v2-L-robust},
but includes several different languages.
Interestingly,
\emph{w2v2-L} did not learn to take sentiment into account,
even though it was trained only on English data,
and the results for \emph{w2v2-L-robust} show
that the w2v2 architecture is able to learn sentiment.

%As the tests include testing for usage of linguistic information,
%they show that \emph{w2v2-L-vox} might
%be a better suited model,
%even though \emph{w2v2-L-robust}
%and \emph{hubert-L}
%are better in standard accuracy benchmarks.

The average test results indicate
a $19$\% lower robustness for \emph{w2v2-L-xls-r}
compared to \emph{w2v2-L-vox}.
While both models include
VoxPopuli~\citep{voxpopuli2021} in the pre-training data,
\emph{w2v2-L-xls-r} also includes other datasets, 
such as BABEL~\citep{babel2014},
which contains 17 African and Asian languages
and which is the only dataset
in the pre-training with telephone speech.
This imbalance in recording conditions between languages 
may have led to a decrease
in robustness for \emph{w2v2-L-xls-r}.
Whereas for \emph{w2v2-L-robust},
a model with only English pre-training data,
the datasets containing telephone speech
seem to have helped it
reach the highest number of passed robustness tests.

An advantage of tests
in addition to benchmarks
is that they are better at characterising model behaviour,
and allow the exclusion of a model
from application
until it passes certain tests
that might be critical
for the application.
Even though \emph{w2v2-L-robust}
is the best model regarding robustness,
it fails several robustness tests,
that add background noise
or use different recording conditions.
Hence,
it might not be suited
for real world applications
without further augmentations
during pre-training,
fine-tuning,
or knowledge distillation~\cite{hinton2015}.

%%%%%%%%%%%%%%%%%%%%%%%%%%%%%%%%%%%%%%%%%%%%%%%%%%%%%%%%%%%%%%%%%%%%%%%%%%%
\section{Conclusion}

%BS: conclusion should usually be in past tense - I changed.

% Number of tests:
% https://gitlab.audeering.com/aderington/ser-tests-paper/-/merge_requests/23
We proposed a large set of 2,029 different tests
to judge the behaviour of speech emotion recognition models
in terms of correctness, fairness, and robustness
on the tasks of predicting
arousal,
dominance,
valence,
and categorical emotions like anger.
The tests allow
to request a certain amount of correctness
or robustness
depending on the desired application of the models.
We further provided an approach
to estimate test thresholds automatically
for testing model fairness.

When applying the test suite
to a selection of eleven models
all trained on \acs{msppodcast},
the results show
that the number of overall passed tests
of the models
relates to rankings on classical speech emotion recognition benchmarks
based on \ac{CCC} and \ac{UAR},
but also shows important differences
and highlights potential shortcuts a model might take.
For example, the four best performing models
in terms of accuracy,
were relying on sentiment of the spoken English text
to boost their valence results.
Whereas the model, that topped the test ranking,
showed lower correctness for valence.

The test results also show
that most models have slight biases
for language,
and that they are not robust enough
for applications
that involve different microphones
or background noise.
Hence,
the tests directly indicate
how to further improve those models.

%%%%%%%%%%%%%%%%%%%%%%%%%%%%%%%%%%%%%%%%%%%%%%%%%%%%%%%%%%%%%%%%%%%%%%%%%%%
\section{Acknowledgements}
The authors would like to thank
A. Triantafyllopoulos,
C. Oates,
and A. Hvelplund
for their valuable discussions and feedback
during the development of the tests,
and J. Wagner for training some of the models
tested in this paper.

%%%%%%%%%%%%%%%%%%%%%%%%%%%%%%%%%%%%%%%%%%%%%%%%%%%%%%%%%%%%%%%%%%%%%%%%%%%
\section{\refname}
 
\printbibliography[heading=none]

\end{document}